\documentclass[]{aa}
\usepackage{epsfig}
\usepackage{natbib}
\usepackage{graphics}
\usepackage{graphicx}
\usepackage{multirow}
\usepackage{amssymb}
\usepackage{color}
\usepackage{boxedminipage}
\usepackage{float}
\usepackage{amsmath}
\usepackage{times}
\usepackage[varg]{txfonts}
\usepackage{verbatim} 
\usepackage{multirow,bigdelim} 
\usepackage{vmargin}
\setmarginsrb{1.2cm}{1cm}{1.2cm}{1cm}{1cm}{1cm}{1cm}{1cm}

\newcommand{\hmsun}{{\,\rm h^{-1}M}_\odot}

  \makeatletter
    \renewcommand{\paragraph}{\@startsection{paragraph}{4}{\z@}%
      {-3.25ex\@plus -1ex \@minus -.2ex}%
      {1.5ex \@plus .2ex}%
      {\normalfont\small\centering}}
     
    \renewcommand{\subparagraph}{\@startsection{subparagraph}{5}{\z@}%
      {-3.25ex\@plus -1ex \@minus -.2ex}%
      {1.5ex \@plus .2ex}%
      {\normalfont\small\centering}}
    \makeatother

\newcommand{\ginnungagap}{{\sc Ginnungagap}}

\newcommand{\gadget}{{\sc Gadget}}

\newcommand{\hMpc}{{ h$^{-1}$~Mpc}}

\begin{document}

\title{Merger types forming the Virgo cluster in recent gigayears}

 \author{Mark Olchanski\inst{1}\thanks{marko@physik.hu-berlin.de}
 \and Jenny G. Sorce\inst{2,3,4}\thanks{E-mail:
 \text{jenny.sorce@univ-lyon1.fr / jsorce@aip.de}}}
\institute{Humboldt Universit\"at zu Berlin - Institut f\"ur Physik, 10099 Berlin, Germany
\and Universit\'e de Strasbourg, CNRS, Observatoire astronomique de 
Strasbourg, UMR 7550, F-67000 Strasbourg, France
\and Leibniz-Institut f\"{u}r Astrophysik, An der Sternwarte 16, 14482 Potsdam, Germany
\and Univ Lyon, Univ Lyon1, Ens de Lyon, CNRS, Centre de Recherche Astrophysique de Lyon UMR5574, F-69230, Saint-Genis-Laval, France\\}


\abstract{As our closest cluster-neighbor, the Virgo cluster of galaxies is intensely studied by observers to unravel the mysteries of galaxy evolution within clusters. At this stage, cosmological numerical simulations of the cluster are useful to efficiently test theories and calibrate model. However, it is not trivial to select the perfect simulacrum of the Virgo cluster to fairly compare in detail its observed and simulated galaxy populations that are affected by the type and history of the cluster.}
{Determining precisely the properties of Virgo for a later selection of simulated clusters becomes essential. It is still not clear how to access some of these properties such as the past history of the Virgo cluster from current observations. Therefore, directly producing effective simulacra of the Virgo cluster is inevitable.}  
{Efficient simulacra of the Virgo cluster can be obtained via simulations that resemble the local Universe down to the cluster scale. In such simulations, Virgo-like halos form in the proper local environment and permit assessing the most probable formation history of the cluster. Studies based on these simulations have already revealed that the Virgo cluster has had a quiet merging history over the last seven gigayears and that the cluster accretes matter along a preferential direction. }
{This paper reveals that in addition such Virgo halos have had on average only one merger larger than about a tenth of their mass at redshift zero within the last four gigayears. This second branch (by opposition to main branch) formed in a given sub-region and merged recently (within the last gigayear). These properties are not shared with a set of random halos within the same mass range. }
{This study extends the validity of the scheme used to produce the Virgo simulacra down to the largest sub-halos of the Virgo cluster. It opens up great prospects for detailed comparisons with observations, including substructures and markers of past history, to be conducted with a large sample of high resolution ``Virgos'' and including baryons,  in the near future. }


\maketitle

\section{Introduction}

The Virgo cluster of galaxies is our closest cluster neighbor. As such it receives much-in-depth attention from observers aiming to understand galaxy formation and evolution within clusters \citep[e.g.][for a non-extensive list]{2000eaa..bookE1822B,2009eimw.confE..66W,2011MNRAS.416.1996R,2012A&A...543A..33V,2011MNRAS.416.1983R,2011ASPC..446...77F,2012ApJS..200....4F,2012MNRAS.423..787T,2014MNRAS.442.2826C,2014ApJ...782....4K,2014A&A...570A..69B,2015A&A...573A.129P,2016ApJ...823...73L,2016ApJ...824...10F,2016A&A...585A...2B}. However, from the numerical side, it is a real challenge to obtain a good simulacrum of the Virgo cluster to precisely compare the simulated and observed galaxy populations to test and calibrate galaxy formation and evolution models. The parameters that the numerical cluster should reproduce to be considered as an efficient simulacrum of the Virgo cluster are simply difficult to completely determine. The formation history of the cluster is of particular importance since optimal comparisons between observed and simulated galaxy populations imply that observed and numerical clusters should have formed from similar mass subhalos at the time of merging \citep[e.g.][for the stellar-to-halo mass ratio]{2015ApJ...807...88G}.\\
Simulations that resemble the local Universe are an interesting approach to determine the detailed formation history of the Virgo cluster and even to obtain high quality Virgo simulacra \citep[e.g.][]{1987ApJ...323L.103B,2001ApJS..137....1B,2010arXiv1005.2687G,2010MNRAS.406.1007L,2013MNRAS.429L..84K}. These simulations stem from initial conditions that have been constrained with observational data that are either radial peculiar velocities \citep[e.g.][]{2008ApJ...676..184T,2013AJ....146...86T,2016AJ....152...50T} or redshift surveys \citep[e.g.][]{2011MNRAS.416.2840L,2012ApJS..199...26H}. Different techniques permit reconstructing the constrained initial conditions either forward \citep[e.g.][]{2013MNRAS.432..894J,2013MNRAS.435.2065H,2014ApJ...794...94W} or backward \citep[e.g.][]{1989ApJ...336L...5B,1990ApJ...364..349D,1991ApJ...380L...5H,1992ApJ...384..448H,1996MNRAS.281...84V,2008MNRAS.389..497K,2008PhyD..237.2139L,2016MNRAS.457..172L}. Resulting simulations reproduce the local Large Scale Structure as well as smaller structures down to the cluster scale \citep[e.g.][]{2016MNRAS.455.2078S}. These simulations have then the merit of reproducing the environment of the Virgo cluster and the cluster itself in its entirety nowadays. In addition, \citet{2016MNRAS.460.2015S} showed recently that the resulting Virgo halos not only share a similar quiet merging history within the last seven gigayears but also that they form along a preferential direction in agreement with the theoretical formation history of the Virgo cluster established from observations \citep{2000ApJ...543L..27W}.\\

In this paper, the merger trees of the Virgo-like halos are studied in more details. Namely, while in the previous study the main focus was onto the Virgo cluster in its entirety (i.e. all the particles that constitute the cluster at z=0), in this paper we have extended the work to the merger trees and their branches to determine how deeply the simulations are indeed constrained. In other words, we seek to quantify up to what level the merger tree scatter expected from random halos within the same mass range as the Virgo cluster is reduced for the Virgo halos, but also how efficient simulacra they can be for further studies of substructures and galaxy populations. The paper opens with a short description of the 15 constrained simulations used for the study proposed in this paper and of the unique Virgo candidate identified in each one. In a third section, the merger trees and in particular the main and second (second after the main) branches are studied in detail. Finally, we conclude that the constrained scheme proves to be efficient to some extent also at the merger tree level and provides some further indications regarding the mergers that formed the Virgo cluster within the last few gigayears.


\section{Virgo halos}

\subsection{Constrained simulations}

\citet{2016MNRAS.455.2078S} described in detail the scheme used to build the constrained initial conditions and to run the simulations. Furthermore the introduction of this paper summarises the various existing techniques. Thus, we summarise in this section only the main steps required to produce the simulations constrained with observational radial peculiar velocity catalogs we use here. We also give a brief description of their purpose with the latest references in the literature of the algorithms used:
\begin{itemize}
\item grouping \citep[e.g.][]{2015AJ....149..171T,2015AJ....149...54T} of the radial peculiar velocity catalog to remove non-linear virial motions that would affect the linear reconstruction obtained with the linear method \citep[e.g.][]{2017MNRAS.468.1812S,2017arXiv170503020S}.
\item minimizing the biases \citep{2015MNRAS.450.2644S} inherent to any observational radial peculiar velocity catalog.
\item reconstructing the cosmic displacement field with the Wiener-Filter technique \citep[linear minimum variance estimator, in abridged form WF,][]{1995ApJ...449..446Z,1999ApJ...520..413Z} applied to the peculiar velocity constraints.
\item relocating constraints to the positions of their progenitors using the Reverse Zel'dovich Approximation and the reconstructed cosmic displacement field \citep{2013MNRAS.430..888D} and replacing noisy radial peculiar velocities by their WF 3D reconstructions \citep{2014MNRAS.437.3586S}. This ensures that, after evolving the structures with an N-body code from an early redshift until today, structures are at the same position (within the 2 \hMpc\ limit of the linear-threshold) to that observed. Keeping constraints at their current position to build initial conditions would indeed result in a shift of some $\ge $10 \hMpc\ between observed and simulated structures after a complete evolution of the initial conditions until today.
\item producing density fields constrained by the modified observational peculiar velocities combined with a random realization to restore statistically the missing structures using the Constrained Realization technique \citep[CR,][]{1991ApJ...380L...5H,1992ApJ...384..448H} 
\item rescaling the density fields to build constrained initial conditions and increasing the resolution by adding small scale features (e.g. \ginnungagap\ code\footnote{https://github.com/ginnungagapgroup/ginnungagap}).
\end{itemize}

\begin{figure}
\includegraphics[width=0.5 \textwidth]{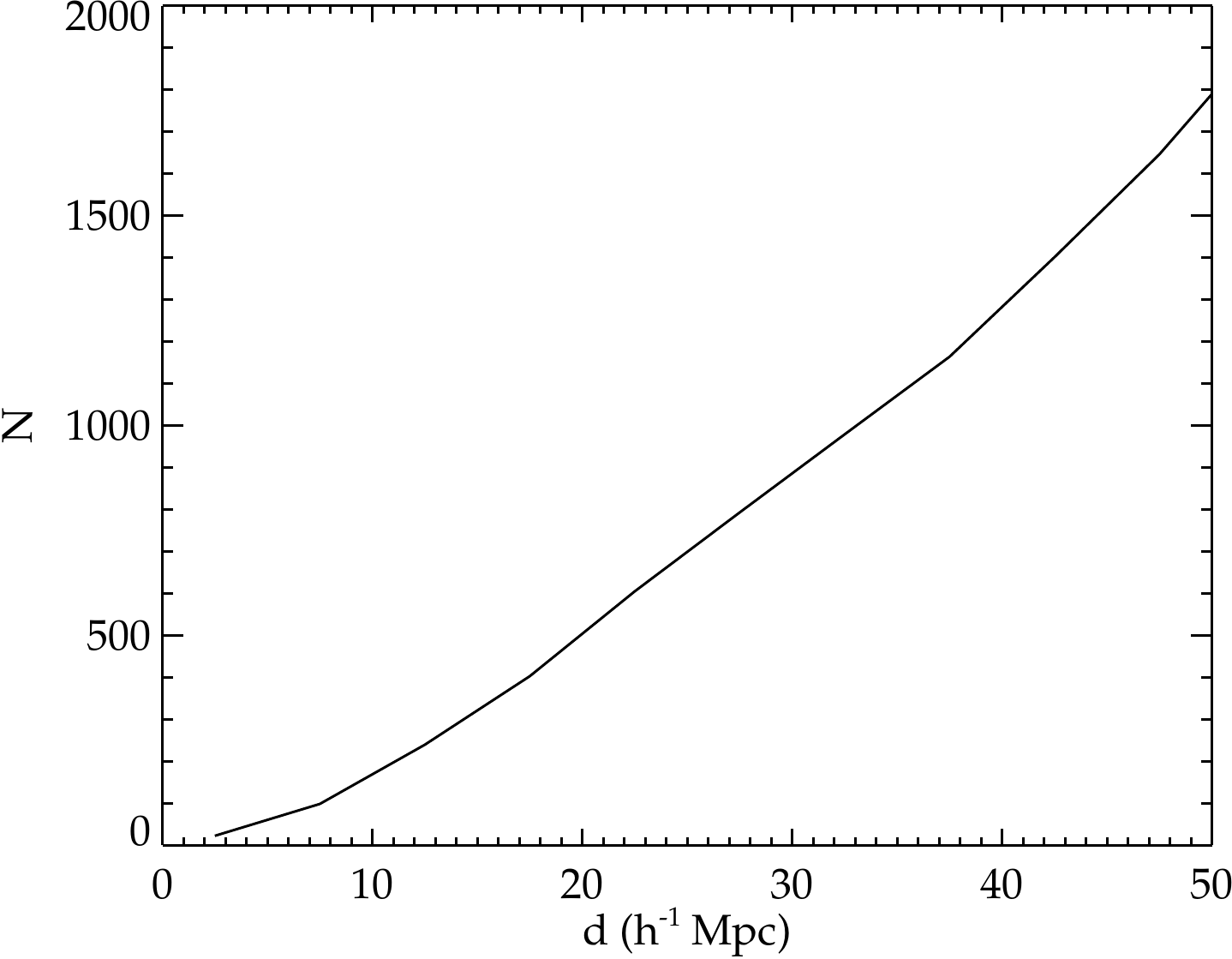}
\caption{Cumulative distribution function of the number of constraints as a function of the distance to the Virgo cluster}.
\label{fig:cumuldistance}
\end{figure}

\citet{2013AJ....146...86T} supplied the observational catalog used as constraints. Since the Virgo cluster is the object of study here, Figure \ref{fig:cumuldistance} permits grasping the distribution of the constraints from the grouped catalog around the Virgo cluster. It presents the cumulative distribution function of the number of constraints as a function of the distance to the Virgo cluster. Initial conditions are evolved within the Planck cosmology framework \citep[$\Omega_m$=0.307, $\Omega_\Lambda$=0.693, H$_0$=67.77, $\sigma_8$~=~0.829,][]{2014A&A...571A..16P} in 500 \hMpc\ boxes with 512$^3$ particles (particle mass: 8$\times$10$^{10}$~$\hmsun$) with the N-body code \gadget\ \citep[][]{2005MNRAS.364.1105S}.

\subsection{Virgo and random halos}

\begin{figure*}
\includegraphics[width=0.95 \textwidth]{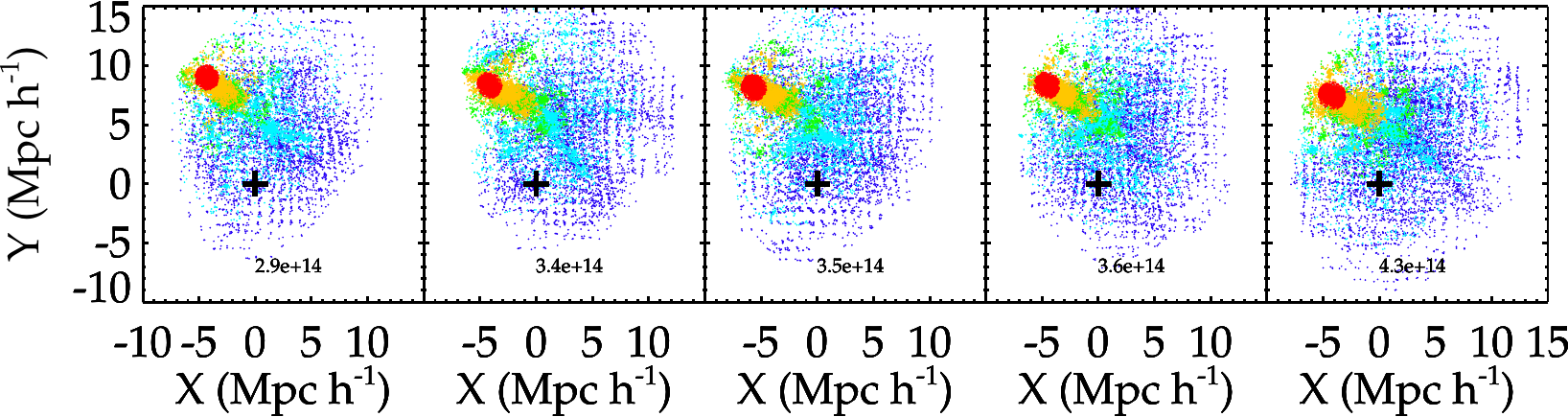} \\
\vspace{0.2cm}

\hspace{-0.1cm}\includegraphics[width=0.95 \textwidth]{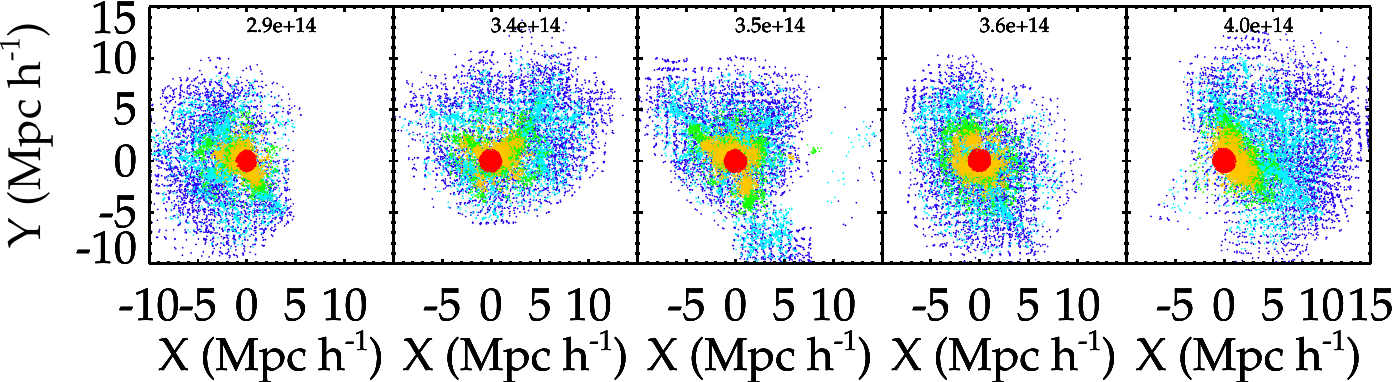}
\caption{Particles belonging to the Virgo simulacra (top) and to random halos (bottom) at redshift zero are shown at different redshifts: 5, 2, 0.5, 0.25 and 0 from dark blue to red. While the position of the Virgo halos is constrained that of random halos is not thus the latter are relocated at (0,0,0) while for the former a cross indicates the center of the box. The gathering of particles forming the Virgo halos at redshift zero is extremely similar for every halo. On the opposite, although they have similar masses, the random halos present aggregation histories along various orientations}. Masses are given in $\hmsun$ for each halo.
\label{fig:appetizer}
\end{figure*}

Subsequently, following \citet{2016MNRAS.460.2015S}, the Virgo dark matter halo in each constrained simulation is identified using the Amiga's Halo Finder \citep[AHF,][]{2009ApJS..182..608K}. By definition of a constrained simulation, the resulting ``Virgos'' are at the proper location with respect to the observer (assumed to be at the center of the box, the box being oriented in the same direction as the local Universe in supergalactic coordinates) and more importantly in the proper local (large scale) environment. 

To quantify the efficiency of the constraining scheme used to produce the local Universe-like simulations, we select a set of random halos that are within the same mass range as the Virgo halos. These random halos are extracted from the constrained simulations. To ensure that these halos are not constrained halos or at least that the constrained nature of the simulation does not impact the results, halos are selected successively randomly in the entire simulations and then outside and in the constrained zone of the simulations. Conclusions are identical in the three cases. The observational catalog extends indeed up to about 150~\hMpc\ with 50\% of the data within 60~\hMpc. It is thus completely reasonable to assume that beyond 200~\hMpc, halos are not affected by the constraints while within they are \citep{2016MNRAS.455.2078S}.

For further studies, 15 Virgo halos and several sets of 15 random halos are thus at our disposal. There are two ways to determine whether a property is constrained  by the constraining scheme: 1) the mean value of the property differs between random and Virgo halos, 2) the range of possible values for the property is smaller for Virgos than for random halos. In other words, the standard deviation of the property is smaller for the former than for the latter. A property can fulfill both conditions, that is, not only the mean but also the standard deviation differ significantly between Virgo and random halos with a smaller standard deviation in the former case that in the latter case.

Figure \ref{fig:appetizer} acts as a summary of previous studies of the Virgo simulacra \citep{2016MNRAS.460.2015S} and as an illustration of the study in the rest of this paper. Its upper panel shows the gathering of particles that belong to five Virgo halos from redshift 5 to redshift 0 through redshifts 2, 0.5 and 0.25 (from dark blue to red). This same gathering is represented in the bottom panel for five random halos with similar masses as the Virgo halos. Since the position of the random halos in the box is arbitrary, the latter are relocated at (0,0,0) while Virgo halos are left at their positions in the box. Virgo halos are extremely similar in terms of positions. Moreover, particles that constitute them at redshift zero gather in the same way and come from the same location in the box. On the contrary, random halos present various formation history at every redshift presented here. Inspecting the YZ and XZ planes reveals the same behavior. These observations to be cumulated with those made by \citet{2016MNRAS.460.2015S} clearly state that choosing random halos within the same mass range as the Virgo cluster are a necessary but not sufficient condition to obtain proper simulacra of the Virgo cluster for further study of its substructure and galaxy population. In the next section, these qualitative observations are quantified using merger trees.

\section{Progenitors}
\begin{figure}
\vspace{-1cm}
\hspace{-0.4cm}\includegraphics[width=0.52 \textwidth]{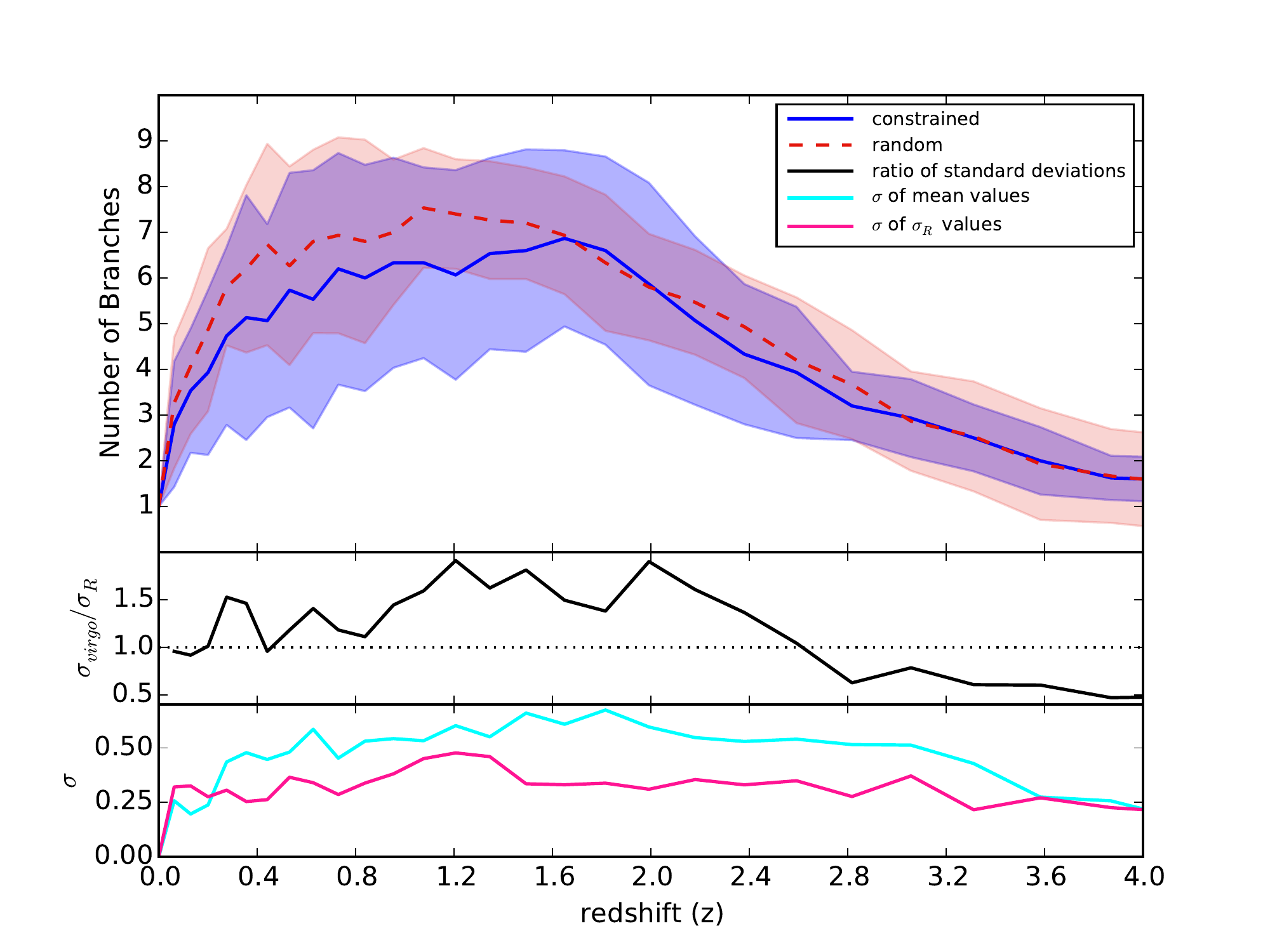} \\

\vspace{-0.5cm}
\hspace{-0.4cm}\includegraphics[width=0.52 \textwidth]{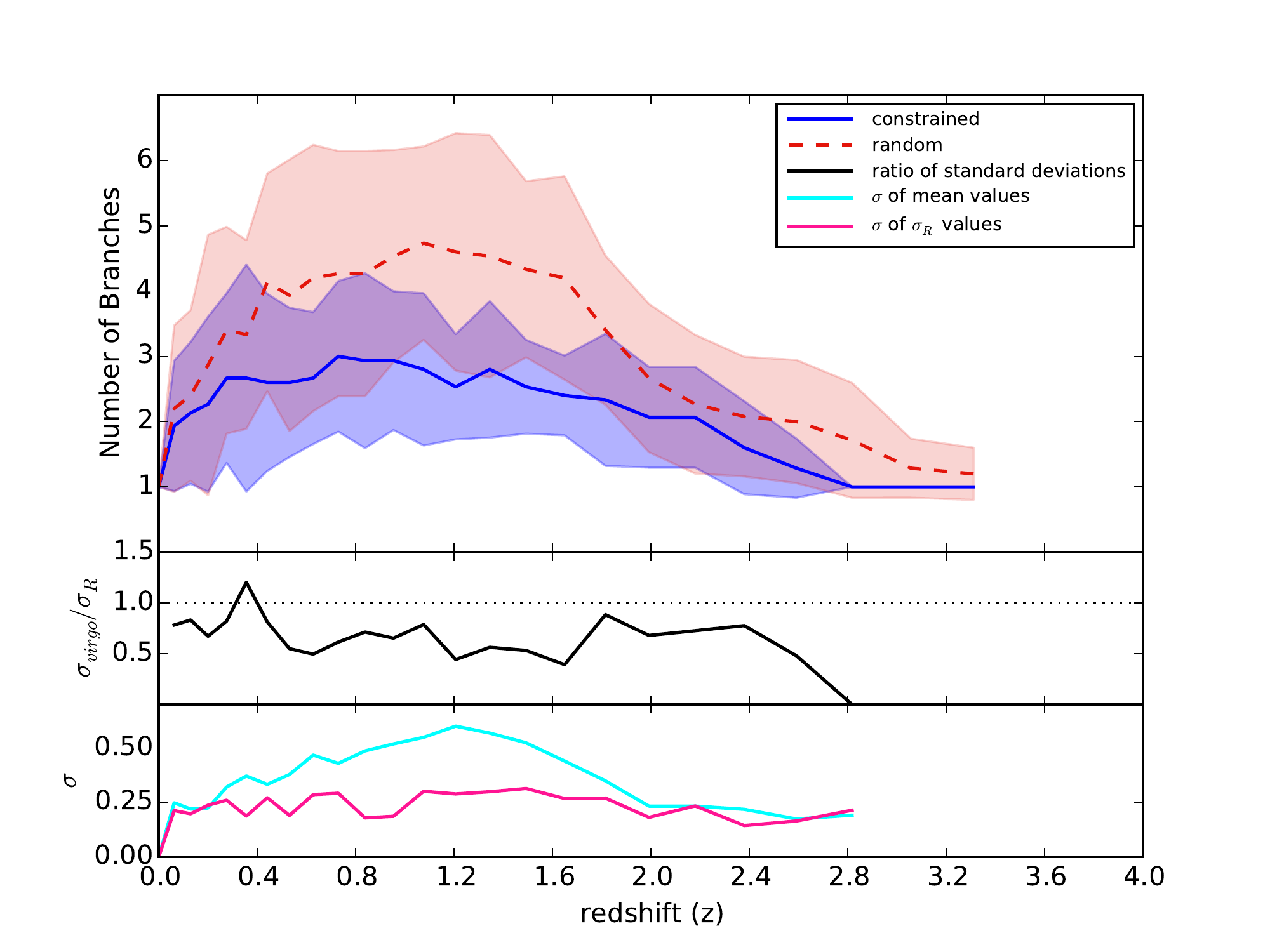}
\caption{First row of both panels: Mean (solid and dashed lines) number of branches and the associated standard deviation (filled areas) for the 15 Virgo (blue) and 15 random (red) candidates. While in the top panel, every halo detected by the halo finder at each redshift is included in the count, in the bottom panel, only halos with more than 100 particles are retained. Results are identical: Virgo candidates have on average a merger tree with less branches than the random ones. Second row of both panels: Ratio (solid black line) of the standard deviations of the constrained to the random samples. Last row of both panels: standard deviation of the means (solid light blue line) and standard deviations (solid pink line) obtained for different sets of 15 random candidates. The low values demonstrate that the results do not depend (small variation of the mean and standard deviation from one sample to the other) on the set of 15 random halos used for the comparison.}
\label{fig:nobranches}
\end{figure}

In this paper, the merger trees of the Virgo halos are under scrutiny. Firstly, we saught to understand whether their merger trees are constrained and differ from those of random halos. Secondly, assuming that the random and constrained merger trees differ, we studied the properties of their different branches separately to draw additional and specific information regarding the formation of the Virgo cluster of galaxies at the sub-halo level. This knowledge is of extreme importance for future studies that will include baryons to compare observed and simulated galaxy populations to finally test and calibrate galaxy formation and evolution models. Galaxy populations are indeed not only sensitive to the large scale environment of the cluster \citep{2014A&A...562A..87E} but also to its formation history in particular its past mergers \citep{2017arXiv170703208D}. A similar formation history at the subhalo level is a requisite to legitimate comparisons between observed and simulated galaxy populations down to the details \citep{{2015ApJ...807...88G}}.

\subsection{Merger Trees}

The halo finder detects halos constituted of 20 particles or more. However, approximately 100 particles ensures a better stability of the halos under study. Netherless, since 1) varying the number of particles required in a progenitor halo to be considered as such or 2) changing the resolution of the simulation obviously affects the number of branches in a merger tree, the key point is that whatever selection criterion is applied, the conclusions must be identical. Figure \ref{fig:nobranches} shows that this is the case: whether all the halos detected by AHF (top panel, first row) or only those with more than 100 particles (bottom panel, first row) are considered, Virgo candidates (blue) have on average fewer branches than random candidates (red) within the last few gigayears (z$<$3). More precisely, they have about 10\% fewer branches in the first case and up to 40\% less branches in the second case for redshifts between 0.4 and 1.6: removing the halos with less than 100 particles strengthens the signal by smoothing out the noise. It suggests that Virgo halos have mostly tiny members in their secondary branches while random halos have more massive progenitors. The smallest members can be studied using higher resolution simulations. In a first step, we are interested in the most massive members. Indeed if the latter do not exhibit signs of being constrained, there is a priori no reason for the smaller members to be constrained. Thus, no further information will be available regarding the Virgo cluster. The resolution of the simulations used here allows us to study the most massive progenitors of the Virgo halos. 

Interestingly, considering only the 100 or more particles halos, not only do Virgo halos have on average fewer branches than the random halos but also the standard deviation of this number of branches is smaller for the former case than for the latter. The second row of the second panel of Figure \ref{fig:nobranches} shows indeed that the constrained to random standard deviation ratio of the number of branches is smaller than one for redshifts higher than 0.4. Specifically, the range of possible number of branches is narrower by up to 50\% for Virgo halos than for random halos. The constraining scheme used to build the look-alike of Virgo affects both the mean and the standard deviation of the number of branches in the merger trees.

Given the fact that only 15 random candidates are used in the two first rows of each panel in Figure \ref{fig:nobranches}, one might wonder whether these conclusions are due to this specific set of 15 random halos. The last row of each panel in Figure \ref{fig:nobranches} ensures that this is not the case: the standard deviation of the means and standard deviations obtained for different sets of 15 random halos have low values. Specifically, the mean (standard deviation) of the number of branches changes by less than about 0.6 (0.25) from one set of 15 random halos to another at the 1-$\sigma$ level. In other words, the difference between the constrained and random mean numbers of branches stays significant whatever random set of 15 random halos is used.

In the rest of this paper, only halos with more than 100 particles are considered and values (mean and standard deviation of the parameter under study) are given up to the redshift where a minimum of three halos out of the fifteen (constrained or random) are still available to derive statistics. 35 random sets of 15 random halos are used to study the impact of the 15 random halos used for comparisons. This number has been selected after checking that increasing it further does not affect anymore the values obtained for the standard deviation of the means and standard deviations.

\subsection{Main \& Second Branches}

An additional individual study of the Virgo halos' merger trees shows that they are overall constituted of one prominent main branch and only one smaller  - but larger than about a tenth of the average mass of the Virgo halos at redshift zero - second branch within the last four gigayears. In addition, all these second branches merge with the main progenitor within the last gigayear. On the contrary,  at the current resolution within the same recent gigayears, random halos can not only have relevant second branches but also third or more branches with more than a few hundred particles. This indicates a first suggestion that the constraining scheme does not only constrain the merging history of the Virgo halos in general \citep{2016MNRAS.460.2015S} but also their merger tree. This low rate of mergers within the last few gigayears is in agreement with observations of the Virgo cluster. More precisely, observations shows that the substructures of the cluster are mostly dominated by massive early-type galaxies. The most massive of these galaxies can only have been formed through major merging events that occurred far in the past. The other galaxies are clearly located in the core of the cluster and their properties indicate that they are completely virialized and thus members since early epochs. As for blue star-forming systems, they are mainly dispersed at the periphery of the cluster suggesting that they started falling recently but independently. They are not part of a major merging event \citep{2014A&A...570A..69B}. \\
Within the last few gigayears, the merger trees of Virgo halos seem to be left with only two prominent branches, the main one and the smaller second one, while they can have many more prominent branches for random halos.  Hereafter, we study in more detail the two branches of the Virgo halos. For the sake of comparisons, the main branch and the most prominent second branch of random halos are also studied. The case of the random halos is a bit more difficult to deal with than that of Virgo halos. While the selection of the second branch is clear for Virgo halos - the only branch within the last 4 gigayears with a mass about a tenth of the average mass of the Virgo halos at redshift zero - in the case of the random halos the second branch to be compared with that of the Virgo halos is harder to define. We select the random second branch as the most massive progenitor after the main one that merged within the last gigayear with the latter. This selection is justified by the fact that it makes more sense to compare structures at the same age, namely structures that had the same time to form and grow.

\begin{figure}
\vspace{-0.6cm}
\hspace{-0.8cm}\includegraphics[width=0.6 \textwidth]{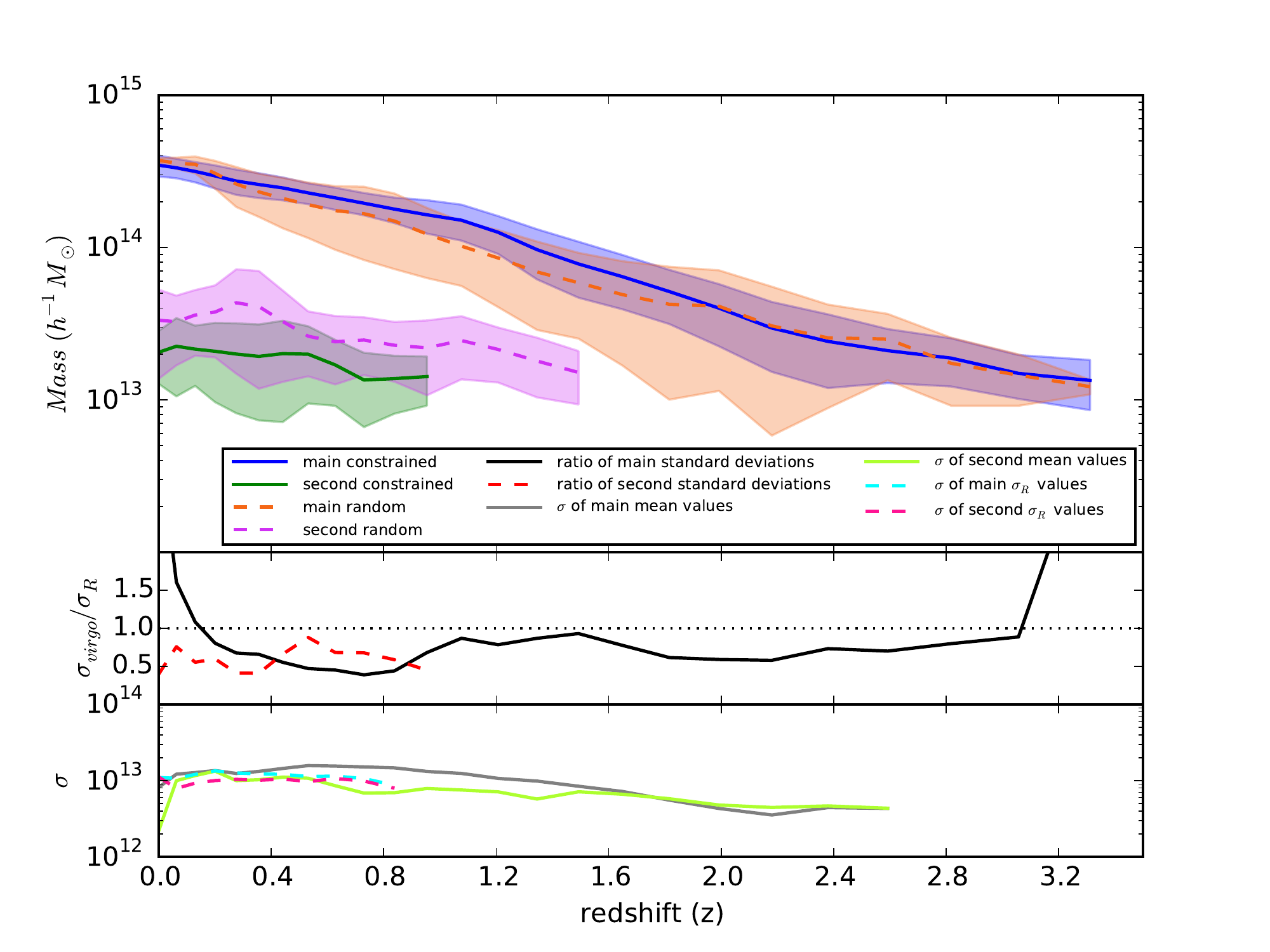} 
\caption{First row: Average merging history of the main progenitor of Virgo (solid blue line) and random (dashed orange line) candidates. Solid green and dashed violet lines stand for the merging history of the second progenitor of Virgo and random candidates respectively before they merge with the main progenitor. With the same color code, the filled areas denote the standard deviations. Second row: Ratio of the constrained to random standard deviations for the main (solid black line) and second (dashed red line) progenitors. Third row: Standard deviation of the means (main: solid gray line and second: solid light green line ) and standard deviations (main: dashed light blue line and second: dashed pink line) of several sets of 15 random candidates.}
\label{fig:mass}
\end{figure}

\citet{2016MNRAS.460.2015S} showed that the Virgo cluster has had a quiet merging history within the last seven gigayears. The main progenitors of the Virgo halos accrete mass at a similar quiet pace and in that respect the main branch of their merger tree is constrained. It is not immediately obvious that once the main branch of the merger tree is somewhat constrained, the second branch is. It is solely because, in addition, the second branch is the only other prominent branch left or in other words, because the whole merger tree (in particular the number of branches) is somewhat constrained. Subsequently, Figure \ref{fig:mass} shows that the mass of the second progenitors is also constrained: the constrained second progenitor (green solid line) is on average smaller than the random one (violet dashed line) by about 2$\sigma$ (2$\times$10$^{13}~\hmsun$).  The last row of the same figure confirms again that this result is independent of the 15 random halos used for the comparison. We note that the second branch does not go down to redshift zero by definition but stops at the latest redshift recorded before z=0 where it still exists. Since the merging happens within the last gigayear, this redshift is very close to zero (about 0.06). Hence the impression that the second branch goes down to redshift zero. 

The second row of Figure \ref{fig:mass} shows that the ratios of the constrained to random standard deviations for both the main (solid black line) and second (red dashed line) progenitors are for most, if not all, redshifts under study here below one (down to 0.5). While for the main progenitor, this result was part of the study of \citet{2016MNRAS.460.2015S}, for the second progenitor, this is a new result: the constraining scheme constrained both the mass of the second progenitor and its range of possible masses ((2$\pm$1)$\times$10$^{13}$~$\hmsun$). We note that this result is only partially due to the quiet merging history of the Virgo cluster within recent gigayears \citep{2016MNRAS.460.2015S}. Indeed, instead of one prominent second progenitor, there could have been lots of small ones merging with the main progenitor in the last gigayear. This is not observed overall for the 15 Virgo halos studied here meaning that the constrained scheme efficiently regulates also the second progenitor of Virgo. \\

\citet{2000ApJ...543L..27W} predicted with observations that the Virgo cluster must have formed along a preferential direction. \citet{2016MNRAS.460.2015S} reinforced this claim since the simulacra of the Virgo cluster formed along a given direction: an auto-correlation function, defined as the distribution of angles formed by two particles infalling onto the cluster and its center of mass divided by the distribution of angles formed by one random point (from an isotropical distribution), one infalling particle and the center of mass, shows clearly a preferential direction of infall. However, this does not necessarily imply that the main and second branches considered separately also `travel' according to their own respective constrained scheme within this region of the box. Consequently, it is interesting to enquire whether within this region of the box, sub-regions can be defined for the different progenitors. To this end, we looked at other parameters such as the velocity components and the displacement or traveled distance of the center of the mass of the progenitor under study (main or second). 

\begin{figure*}
\vspace{1cm}\includegraphics[width=1 \textwidth]{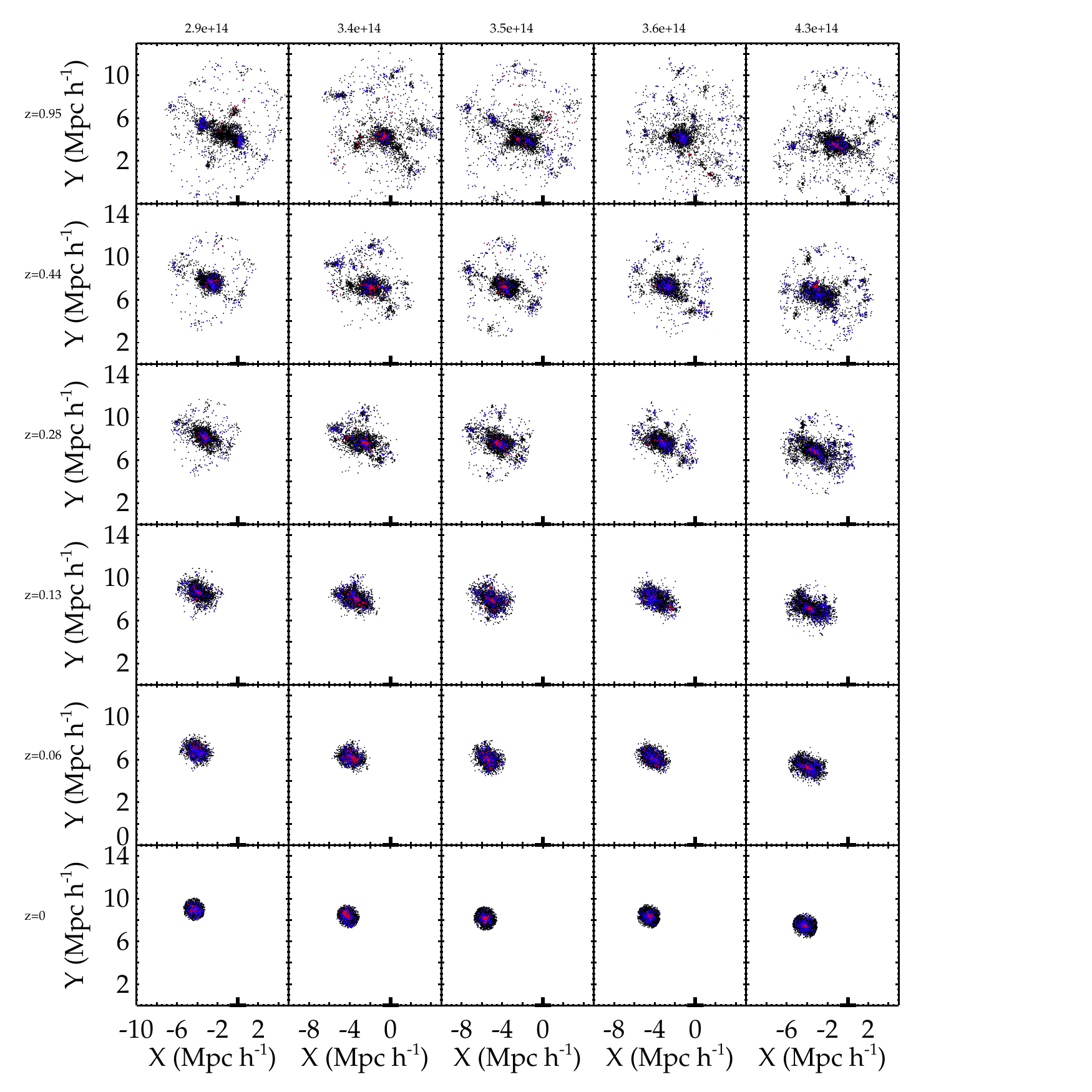} 
\caption{Particles belonging to the main (blue dots) and second (red dots) progenitors at redshift 0.06 are shown at different redshifts (given in the top left of each panel). Black dots indicate the other particles that belong to the halos at z=0. Masses are given in $\hmsun$ for each halo. In the last four gigayears (after z=0.4), there are no mergers between the most massive progenitor and a progenitor with a mass larger than about a tenth of the mass of the halos apart from that with the second progenitor.}
\label{fig:12part}
\end{figure*}

Before this study, Figure \ref{fig:12part} gives a first visual impression of the formation of five Virgo halos within the last seven gigayears. Blue dots represent particles belonging to the most massive progenitor at z=0.06 and the red ones those of the second progenitor that still exists at z=0.06 (it has merged at z=0). Clearly for all the Virgo halos, some mergers greater than a tenth of the mass at redshift zero happen between four and seven gigayears ago (z=0.95 to 0.4) as clumps of blue dots merge onto the main progenitor. However after z=0.4, i.e. within the last four gigayears, the last merger with a mass about a tenth of the mass of the Virgo halos at z=0 that still needs to happen is that with the second progenitor (that is still forming). In addition, for each one of the Virgo halo, blue, red and black dots have a similar motional behavior.

\begin{figure*}
\hspace{-0.8cm}\includegraphics[width=0.54 \textwidth]{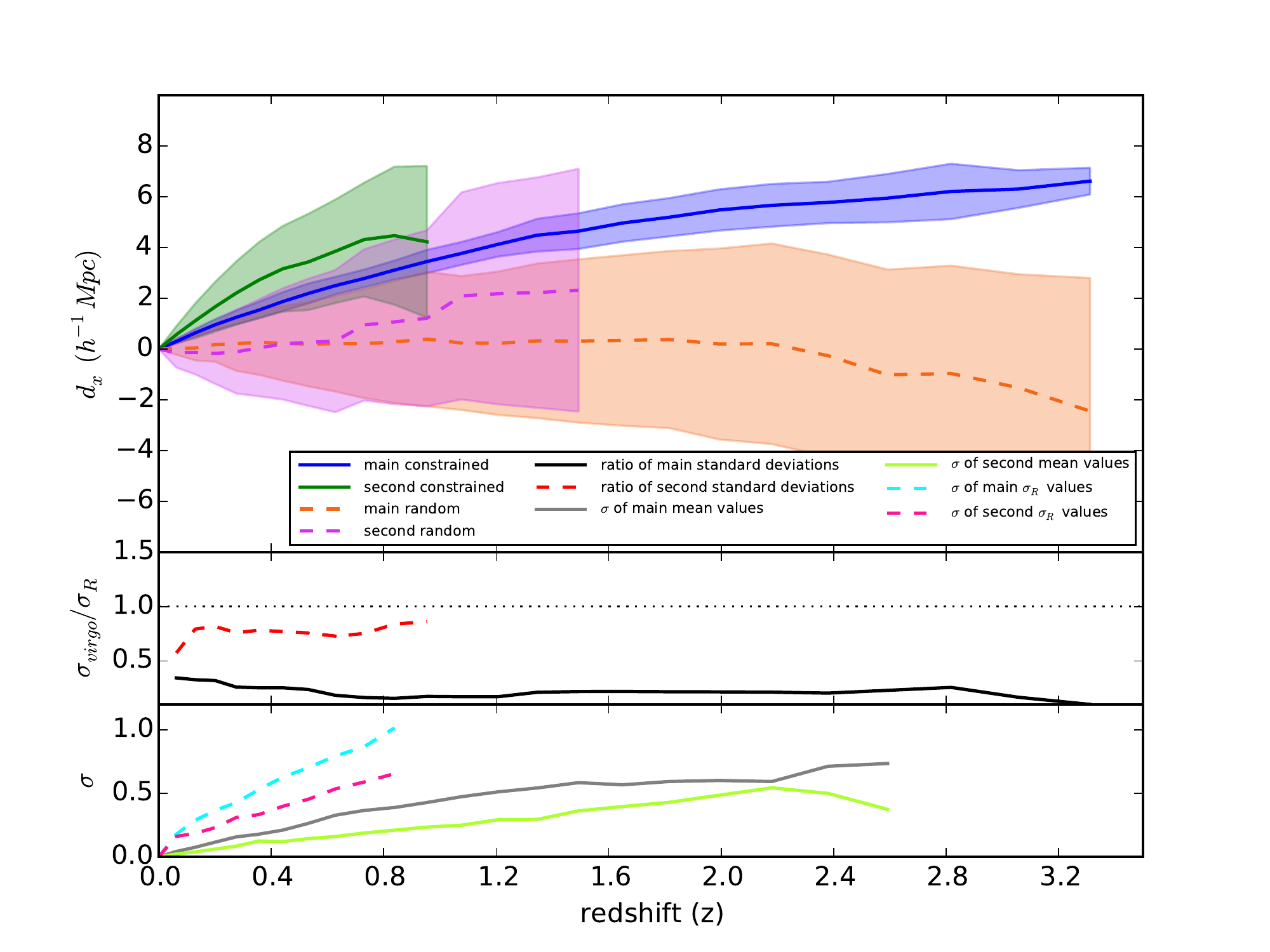} 
\hspace{-0.8cm}\includegraphics[width=0.54 \textwidth]{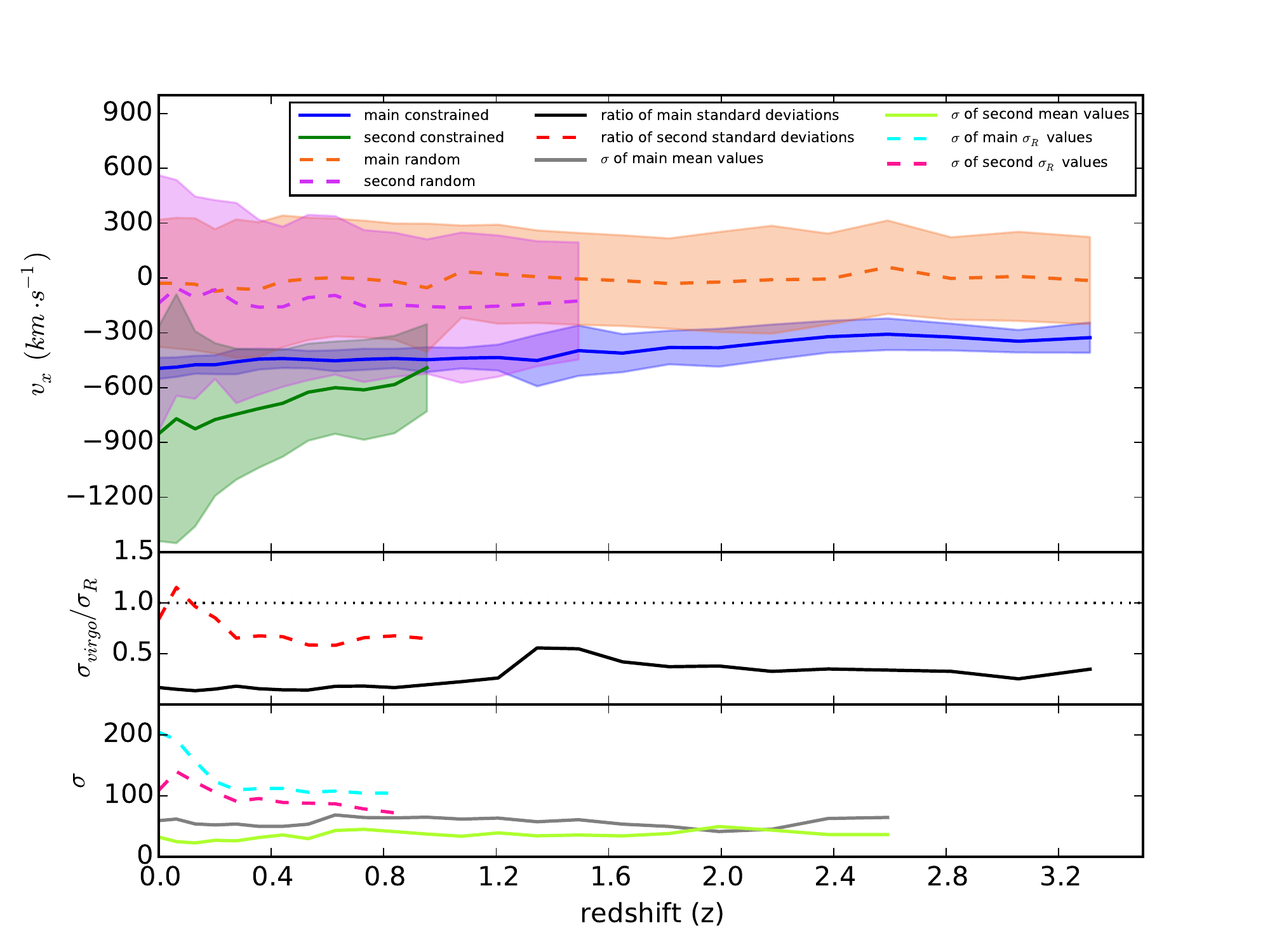} 
\caption{Top row: Average x component of the displacement (right) and velocity (left) of the main and second progenitors of Virgo (solid blue and green lines) and random (dashed orange and violet lines) candidates. The standard deviations are shown with the transparent areas using the same color code. Second row: Ratio of the constrained to random standard deviations for the main (solid black line) and second (dashed red line) progenitors. Bottom row: Standard deviation of the means (main: solid gray line and second: solid light green line) and standard deviations (main: dashed light blue line and second: dashed pink line) of several sets of 15 random candidates.}
\label{fig:dispvel}
\end{figure*}

Figure \ref{fig:dispvel} allows us to investigate these motions more quantitatively for the x component of the velocity and displacement of the main (blue particles) and second (red particles) progenitors. The same color code as in Figure \ref{fig:mass} is used. Appendix A gives similar results for the y and z components. The simulations are oriented in the same way as the observed local Universe and the same supergalactic coordinate system is used. The left (right) panel of Figure \ref{fig:dispvel} shows the x component of the displacement (velocity) of the progenitors with respect to their last recorded redshift of existence (z=0 for the main progenitor, earlier redshift - about 0.06 - for the second progenitor or equivalently the main progenitor at redshift zero in both cases since results are insignificantly different). Results are indisputable: at late times (low redshifts, z$<$0.8) main and second progenitors gather according to their own given way (velocity, displacement) for all the Virgos under study to finally merge and form the Virgo cluster at redshift zero. The effect is the clearest in the x direction as shown on Figure \ref{fig:dispvel} when compared to the y and z directions in Appendix A: the second progenitor travels on average along a given x direction (decrease of the x coordinate in the box oriented to correspond to the x supergalactic coordinate) faster than the main progenitor, probably following the latter and thus being accelerated by it before they both merge at redshift zero. Although the trend is less obvious along the y and z directions as shown in Appendix A, it is not completely inexistent. We note that replacing the reference ``main progenitor at redshift zero'' by ``main progenitor at each redshift'' to compute the displacement does not change the conclusions.

By comparison, Figure \ref{fig:dispvel} shows that the random main and second progenitors display as expected no typical features on average in the x direction (the mean values are close to zero). This assertion is true also for the y and z directions as shown in Appendix A. Again this result does not depend on the set of 15 random halos selected for the comparison: the last row of the panels of Figure \ref{fig:dispvel} shows that the mean and standard deviation vary little from one random set to the other. Additionally, the second row of the panels shows that the possible range of values for the x component of the velocity and displacement of the main progenitor is considerably narrower for the Virgo halos than for the random halos. This is in agreement with the fact that the Virgo halos all accrete matter along a similar direction. The trend is similar when comparing the velocity components of the constrained and random second progenitors. Appendix A shows again that although the signal exists in both the y and z directions, it is clearer in the x direction. The negative x direction being the leading one for the formation of the Virgo cluster and its progenitors considered individually is in complete agreement with the observations that show that the very local Universe (about 15~\hMpc) goes toward the ``Great Attractor'' region located at ``lower x values'' than Virgo \citep[e.g. see][for the earliest references]{1987Sci...237.1296W,1989Natur.338..538K}.

\begin{figure}
\includegraphics[width=0.5 \textwidth]{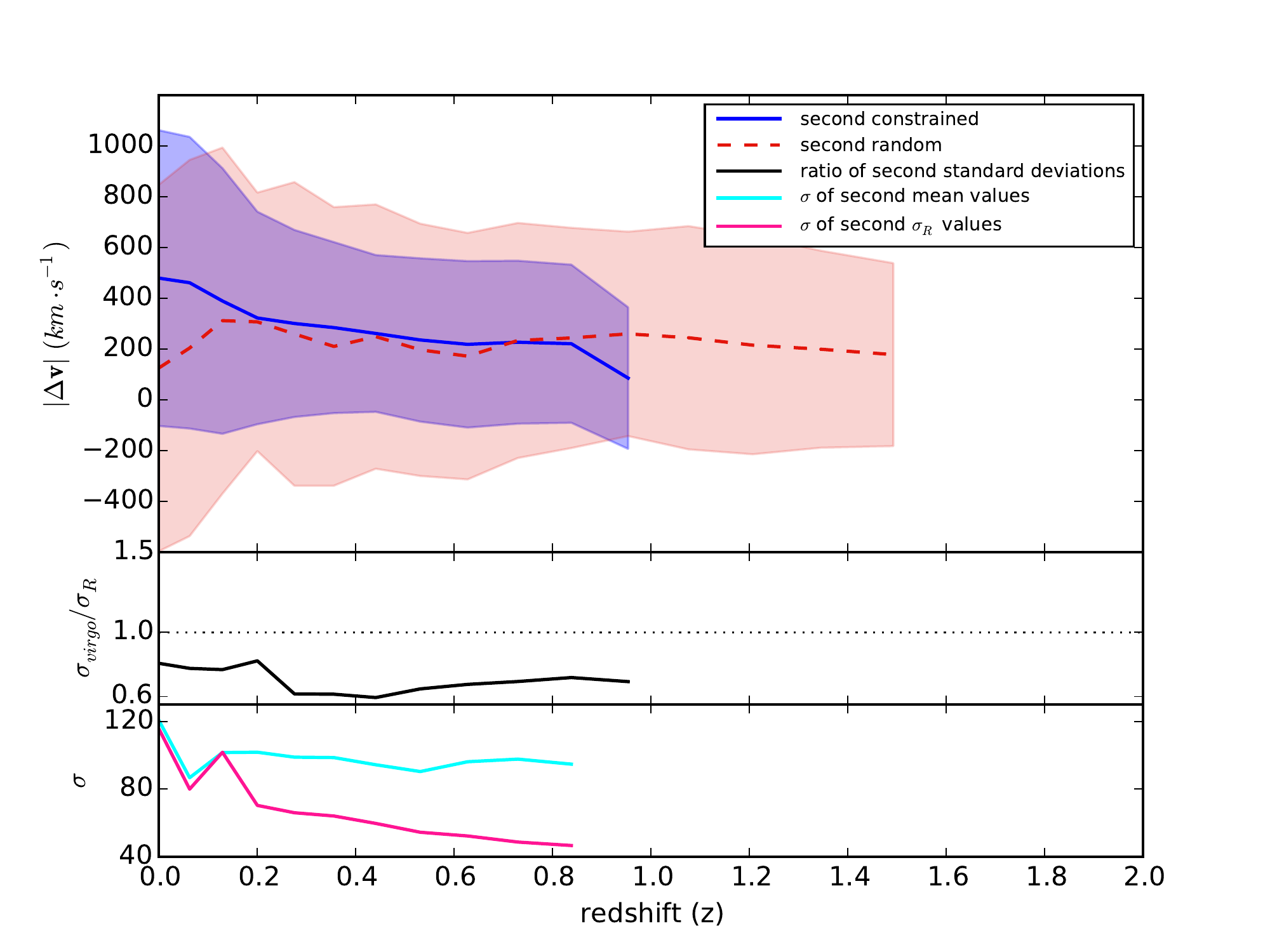}
\caption{Top row: Average relative velocity of the second constrained (blue solid line) and random (red dashed line) progenitors with respect to the main halo at each redshift. Middle row: Ratio of the constrained to random standard deviations of the relative velocity. Bottom row: Standard deviation of the means (solid light blue line) and standard deviations (solid pink line) of several sets of 15 random candidates.}
\label{fig:vel}
\end{figure}

Figure \ref{fig:vel} pushes further the comparisons with the relative velocity of the second progenitors with respect to the main branch at each redshift. Here again, the relative velocity is clearly constrained: the ratio of the constrained to random standard deviations is smaller than one. Additionally, the relative velocity of the constrained second progenitors is on average larger than that of the random second progenitors at redshifts smaller than 0.2.
 
\section{Conclusion}

This paper aims to investigate the constraining power of the scheme - used here to build simulations that look like the local Universe - at the sub-cluster level. It focuses on our closest cluster-neighbor, the Virgo cluster of galaxies. Such a study is important in two linked aspects: 1) it permits gathering information regarding the formation of the Virgo cluster, an essential pre-requisite for later selection of numerical clusters to be able to compare legitimately observed and simulated galaxy populations down to the details ; or alternatively 2) it permits determining the scale down to which the constrained simulated clusters can be used to test and calibrate galaxy formation and evolution models without biases due to variation in the formation history of the simulated clusters with respect to the observed one.

The constrained simulations have already been proven to be excellent reproduction of the local Large Scale Structure \citep{2016MNRAS.455.2078S} as well as of the Virgo cluster as a whole, including its formation history in general \citep{2016MNRAS.460.2015S}. In this paper, we thus further studied the formation histories of the Virgo halos and we focused on their merger trees and in particular on their main and second branches defined hereafter. Comparing the latter to those of random halos within the same mass range, we show that the constraining scheme is still very efficient at these scales, namely at the largest sub-halo scales: 
\begin{itemize}
\item For redshifts between 0.4 and 1.6, the merger trees of Virgo halos have on average 40\% less branches than random ones considering branches with halos more massive than 8$\times$10$^{12}$~$\hmsun$ . This is in agreement with their quiet merging history within recent gigayears.
\item  Within the last four gigayears, Virgo halos had overall, apart from their main progenitor, only one other prominent progenitor in contrast with the random halos that could have several other prominent progenitors. On average, this other prominent progenitor is about a tenth of the mean mass of the Virgo halos at redshift zero and it merged within the last gigayear with the main progenitor. In addition the random second progenitor, defined as the most massive progenitor after the main one that merged within the last gigayear, is 2$\sigma$ more massive than the constrained one. This is in agreement with the quiet history of the Virgo cluster within the last gigayears with the additional information that there was one merger more prominent  (although moderately) than the others that happened within the last gigayear. This numerical statement is in agreement with observations of the galaxy population of the Virgo cluster that imply early major mergers \citep{2014A&A...570A..69B}.
\item At late times, main and second progenitors of Virgo candidates follow their own accretion scheme that appears to be constrained with respect to that of main and second progenitors of random candidates. In particular, Virgo's progenitors exhibit a clear peculiar trend of motion in the negative x supergalactic direction in agreement with our knowledge regarding the motion of the Virgo cluster. The second progenitor moves faster than the main one probably because it is in the wake of the latter. This highlights that the main and second progenitors follow their own accretion scheme in similars way in the different constrained simulations.
\end{itemize}
This paper extends the efficiency of the constraining scheme down to the largest sub-halo level. Besides providing more properties of the Virgo cluster to select optimal simulacra in typical cosmological simulations, it opens great perspectives regarding 1) detailed comparisons with observations, including substructures and markers of the past history visible in the galaxy population, to be conducted in a near future with a large sample of high resolution ``Virgos'', 2) simulations including baryons to test and calibrate galaxy formation and evolution models down to the details.

\section*{Acknowledgements}

The authors thank the referee for very useful comments that helped clarify the paper. JS thanks her collaborators in particular Gustavo Yepes, Stefan Gottl\"ober, Noam Libeskind and Yehuda Hoffman for interesting discussions. The authors gratefully acknowledge the Gauss Centre for Supercomputing e.V. (www.gauss-centre.eu) for providing computing time on the GCS Supercomputers SuperMUC at LRZ Munich and Jureca at JSC Juelich. JS acknowledges support from the Astronomy ESFRI and Research Infrastructure Cluster ASTERICS project, funded by the European Commission under the Horizon 2020 Programme (GA 653477), from the ``Centre National d'\'etudes spatiales (CNES)'' postdoctoral fellowship program as well as from the ``l'Or\'eal-UNESCO Pour les femmes et la Science'' fellowship program.

\appendix\markboth{Appendix}{Appendix}
\renewcommand{\thesection}{\Alph{section}}
\numberwithin{equation}{section}

\begin{appendix}

\section{y and z components of the velocity and the displacement}

\begin{figure*}
\hspace{-0.cm}\includegraphics[width=0.55 \textwidth]{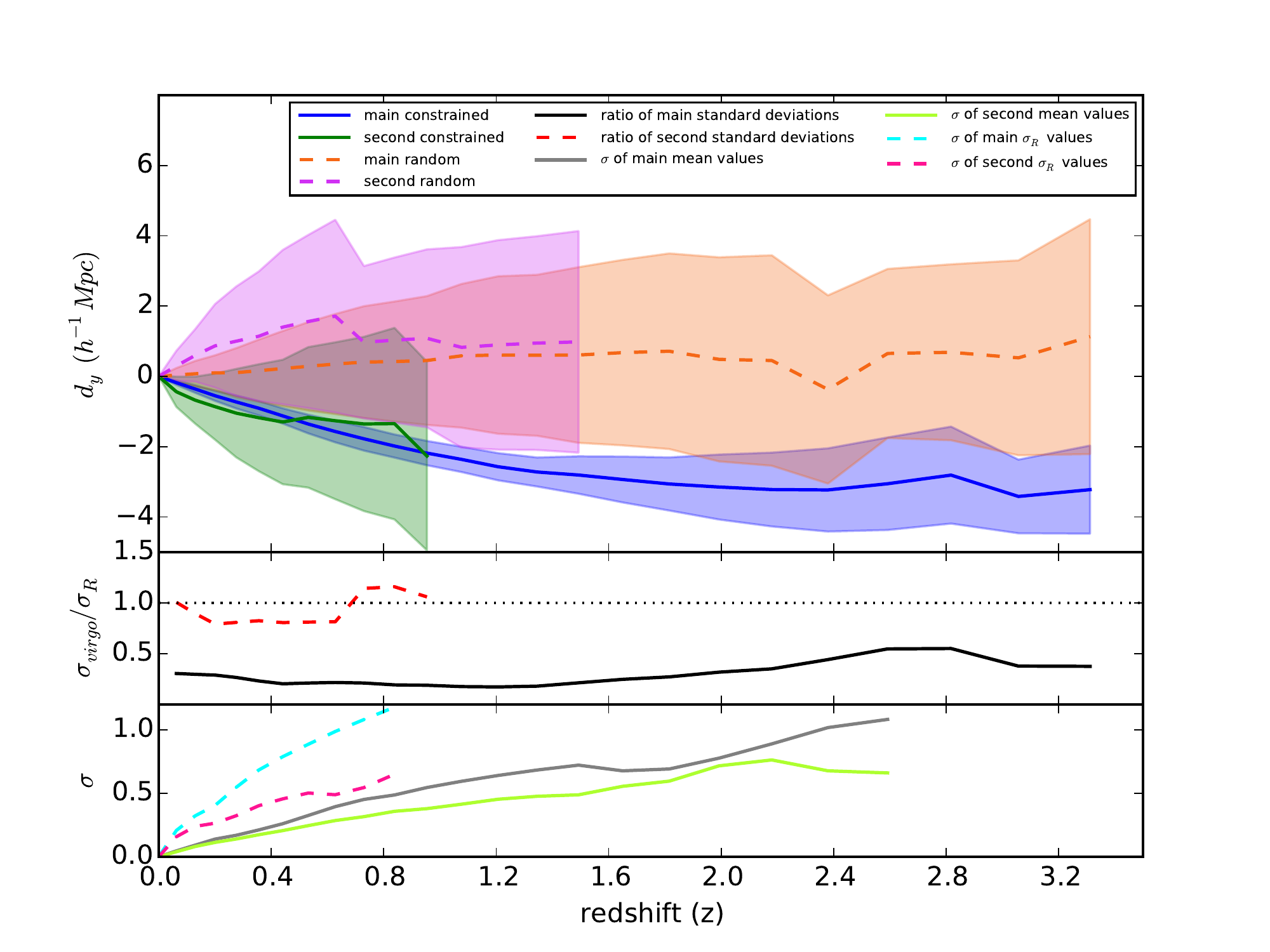} 
\hspace{-0.75cm}\includegraphics[width=0.55 \textwidth]{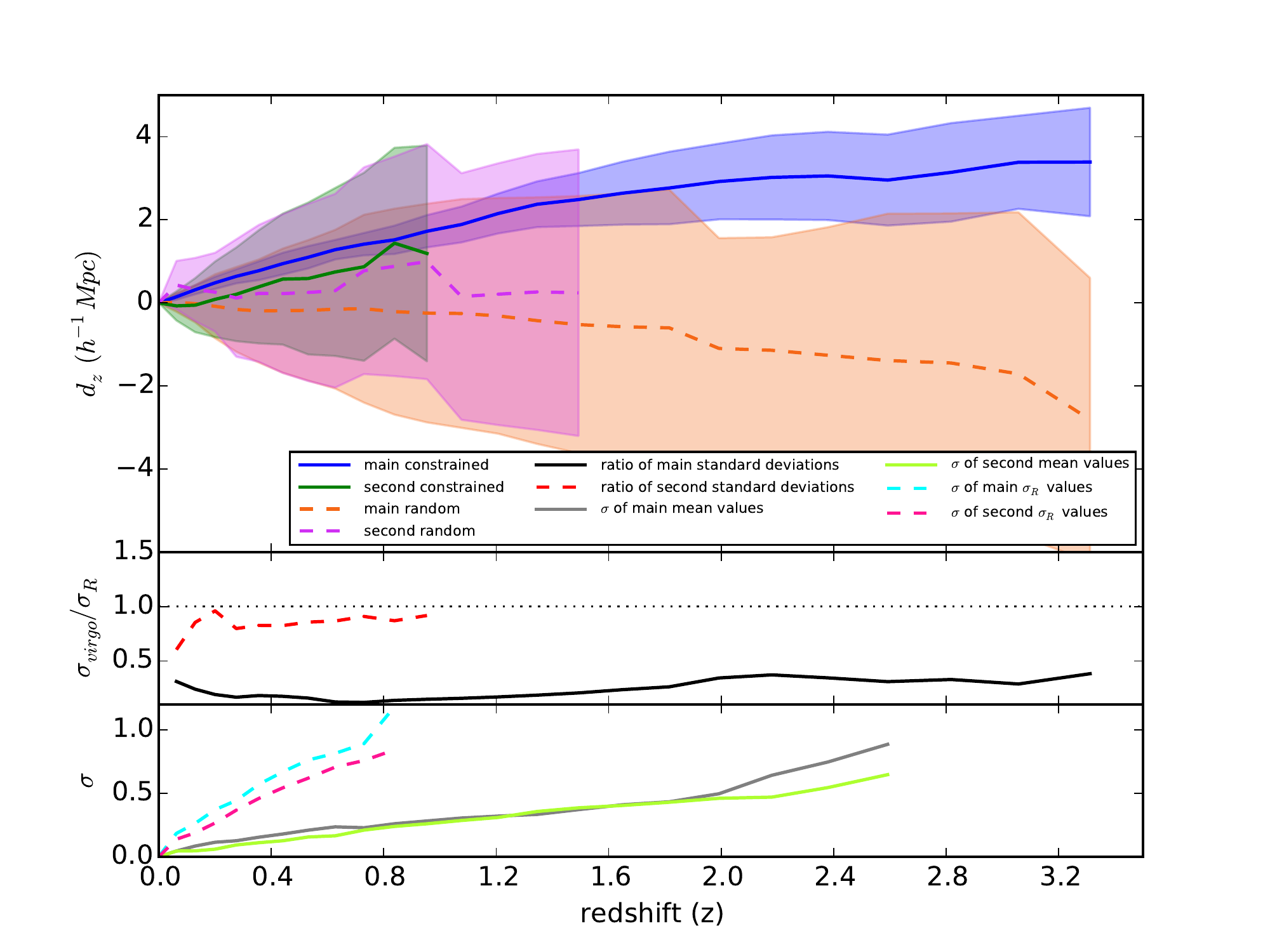} \\
\hspace{-1cm}\includegraphics[width=0.55 \textwidth]{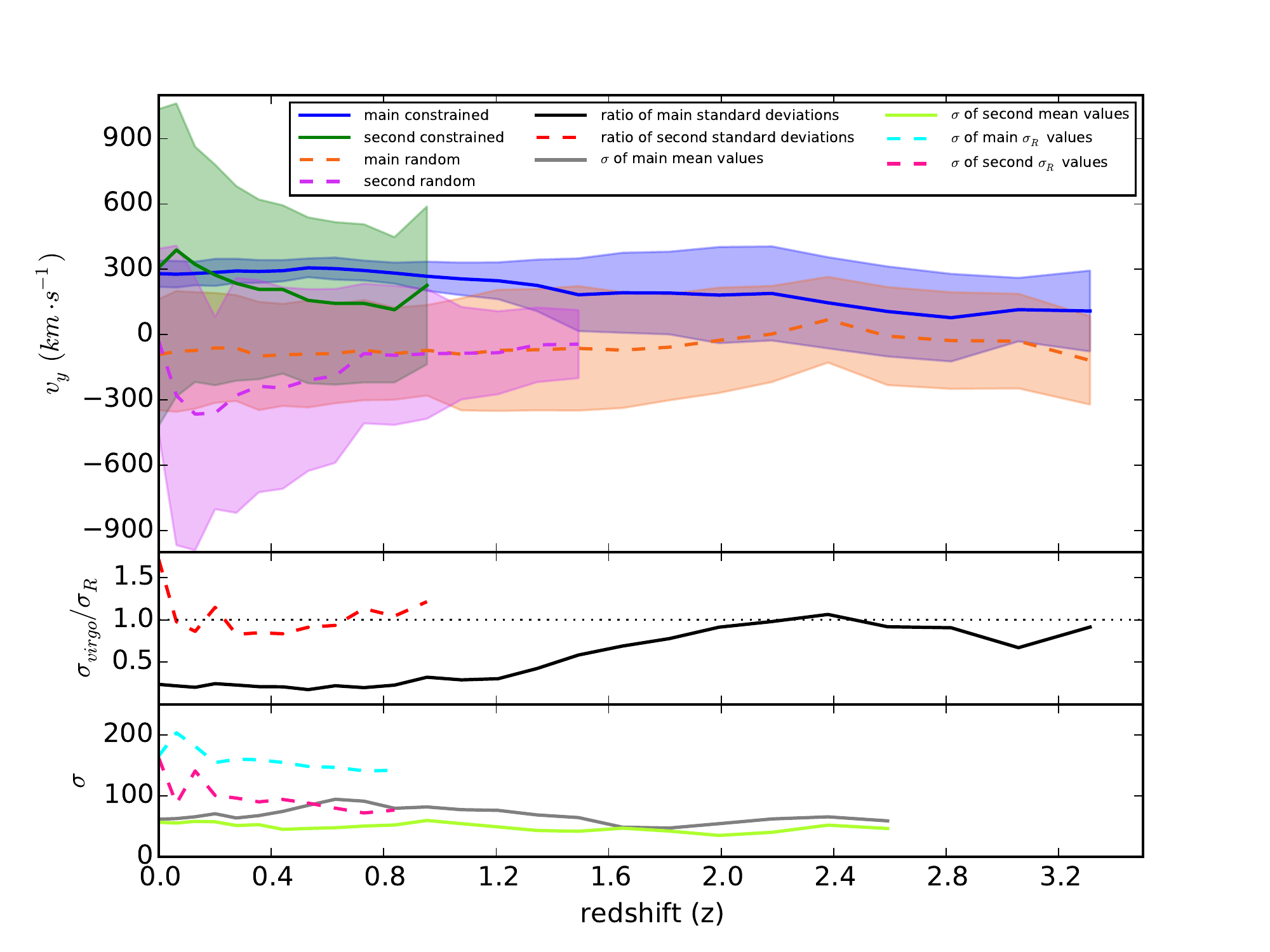} 
\hspace{-0.75cm}\includegraphics[width=0.55 \textwidth]{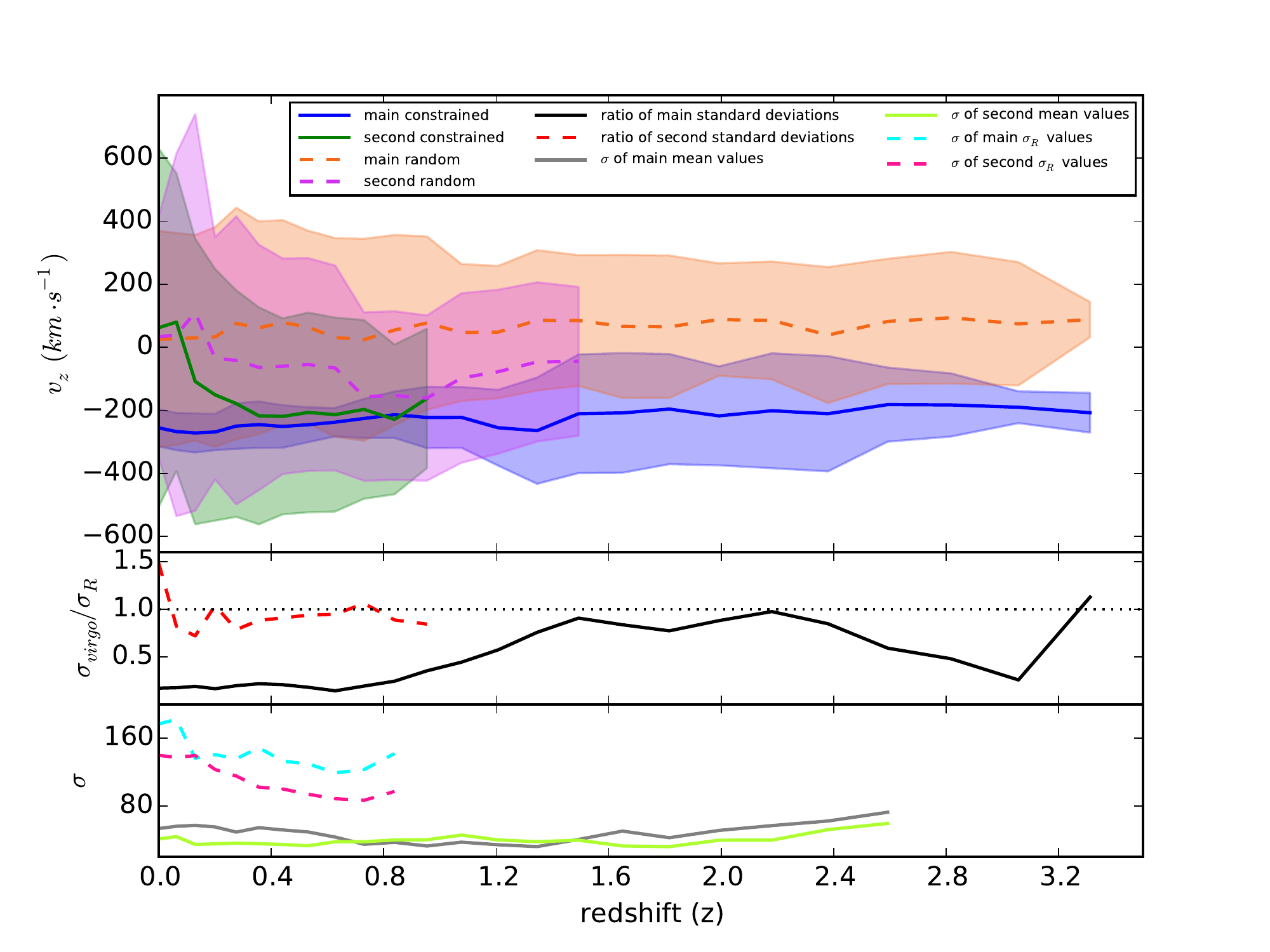} 
\caption{From left to right, top to bottom: Average y and z displacement and velocity (first row of each panel) components of the main and second progenitors of Virgo (solid blue and green lines) and random (dashed orange and violet lines) candidates. The standard deviations are shown with the transparent areas using the same color code. Second row of all the panels: Ratio of the constrained to random standard deviations for the main (solid black line) and second (dashed red line) progenitors. Last row of all the panels: Standard deviation of the means (main: solid gray line and second: solid light green line) and standard deviations (main: dashed light blue line and second: dashed pink line) of several sets of 15 random candidates.}
\label{fig:appendix}
\end{figure*}

Figure \ref{fig:appendix} shows the average y and z components of the displacement and velocity of the main and second progenitors of Virgo and random candidates. The same conclusions as for the x component can be drawn. There exist both a privileged direction of displacement and a privileged velocity value for the main and second branches of Virgo halos that are not found for the branches of random halos as expected.

\end{appendix}
\newpage
\bibliographystyle{aa}

\bibliography{biblicomplete}

\label{lastpage}

\end{document}